\documentclass[12pt,epsfig,epic,eepic]{article}
\usepackage{epsfig,pstricks}
\usepackage{epic}
\usepackage{eepic}

\newcommand{\be}{\begin{equation}}
\newcommand{\ee}{\end{equation}}
\newcommand{\ba}{\begin{eqnarray}}
\newcommand{\ea}{\end{eqnarray}}

\newcommand{\bd}{\begin{description}}
\newcommand{\ed}{\end{description}}

\newcommand{\system}{{\sc system}}

\renewcommand{\iota}{{\bf 1}}

\def\rellow#1#2{\mathrel{\mathop{\kern 0pt #1}\limits_{#2}}}

\def\S{{\bf S}}

\newtheorem{nom}{Nomenclature}

\newtheorem{define}{Definition}

\newtheorem{theorem}{Theorem}

\newtheorem{remark}{Remark}
\title{Life in Silico - Simulation of Complex Systems\\ by Enzymatic Computation}
\author{Gerhard Mack and Jan W\"urthner\\
   II. Institut f\"ur Theoretische Physik, Universit\"at Hamburg}
\date{September 28, 2000}
\begin{document}
\maketitle
$\quad$\\[-5mm]\noindent
{\bf abstract}
We describe software and a language for  quasibiological computations. 
Its theoretical basis is a unified theory of complex (adaptive) systems 
 where all laws are regularities of 
relations between things or agents, and dynamics is made from
 ``atomic constituents'' called enzymes.
The notion is abstracted from biochemistry.
 The software can be used to simulate physical systems 
as well as basic life processes. Systems can be constructed and manipulated 
 by mouse click and 
there is an automatic translation of all operations
into a LISP-like scripting language, so 
that one may compose code by mouse click.  
\section{Introduction}
It is a fruitful idea to learn from nature how to do information processing 
by abstracting from what happens in the living cell.

In 1959, R. Feynman gave a visionary talk describing the possibility of 
building computers that were ``submicroscopic''. \cite{Feynman}.

More recently, L.M. Adleman demonstrated the feasibility of carrying out 
computations
at the molecular level by solving a standard NP-hard graph problem (similar to 
the travelling salesman) using molecules of DNA, and standard protocols and
biochemical enzymes \cite{Adleman}

Here we are not interested in doing real chemistry in a bucket,
 but in mathematical abstractions which incorporate the basic logic of 
biochemical processes in the living cell and can be put on a computer.
 This was  a dream also of workers on Artificial
 Life \cite{artificialLife}. But we may infer from Adleman's work that  {\em life in silico} will be useful not only for simulating life processes, but for other complex problems as well. 

Starting point is a philosophical principle,
\begin{quotation}
The human mind thinks about relations between things or agents. 
\end{quotation}
This tells us how to model parts of the world. 
A unified theory of complex (adaptive) systems 
was built upon this principle \cite{mack:cmp}. 
Here we describe its implementation in software. 

 As in category theory
\cite{CWM}, the relations are directed binary relations, and it is regarded
 as their constitutive property
that they can be composed - think of \emph{friend of a friend, brother in law,
next nearest neighbor}. There are also some distinguished relations -
the identities $\iota_X$ of objects $X$ with themselves - they play a basic
role similar to the number $0$ in arithmetics. Motivated by physics,
a principle of locality is added by singling out some of the 
relations as {\em direct} relations, called {\em links}. All others are 
composed from them and possibly their ``adjoints'' - i.e.
 relations in the opposite direction. 
Links between objects form networks. Their properties are laid down 
in the mathematical definition of a \system , given below. 
Basically it makes precise the notion of {\em structure}. 

We consider dynamics in discrete time. It is required to be local. 
Dynamics is also built from ``atomic constituents'', called enzymes. 
They are composed from very simple basic moves, including in particular 
composition of two links to a single link (friend of a friend becomes friend),
making or deletion of adjoint links (reciprocation) and 
making or finding copies.
 \footnote{They are very simple examples of graph transformations 
\cite{graphGrammar}}

In contrast with automata theory \cite{automata},
 \system s theory is supposed to be self
contained. Everything that is used should be provided for by the axioms
or be constructed with them. There is no other information but 
structure. But in the software implementation, we will not push this 
point of view to the extreme. We will admit numerical or text data
that reside inside objects and links. Regard them as coding for structure. 
For more discussion on the relation between structural and numerical 
descriptions see ref. \cite{mack:cmp}. 

The whole theory may be regarded as a kind of {\em universal chemistry}
where general objects substitute for atoms and molecules, and
more general links for chemical bonds and spacial proximity. 

\section{Systems}
\label{sec:system}
According to the pioneer of general systems theory, L. van Bertalanffy
\cite{Bertalanffy}, 
{\em a system is a set of units with relationships between them.}
And according to F. Jacob \cite{Jacob}, 
{\em every object considered in biology 
is a system of systems}.  We precisize to
\footnote{Axioms 1,3 are those of a category \cite{CWM}. Therefore \system s are 
categories with a $*$-operation and a notion of locality} 
\begin{define}{\em ({\sc System})}
\label{def:system}
A \system \ $\S$ is a 
model of a part of the world as a network of
{\em objects} $X,Y,...$ (which represent things or agents) with {\em arrows} 
$f,g,... $ which represent directed relations between them.\\ One writes 
$f:X\mapsto Y$ for a relation from a {\em source} $X$ to a {\em target}
$Y$.

The arrows are characterized by axiomatic properties as follows:
\begin{enumerate}
\item 
{\em composition}. Arrows can be composed.
If $f:X\mapsto Y$ and $g:Y\mapsto Z$ are arrows,
then the arrow $$g\circ f: X \mapsto Z$$ is defined.
The composition is associative, i.e. $(h\circ g)\circ f= h\circ (g\circ f)$. 
\item
{\em adjoint}.
To every arrow $f: X\mapsto Y $ there is a unique arrow
$f^{\ast }:Y\mapsto X $ in the opposite direction, called the 
adjoint of $f$. $f^{\ast\ast} = f$ and
$(g\circ f)^{\ast}=f^{\ast}\circ g^{\ast}$. 
\item
{\em identity}. To every object $X$ there is a unique arrow 
$\iota_X: X\mapsto X $ which represents the identity of a thing or agent
 with itself. 
$$\iota_X = \iota_X^{\ast}, \qquad \mbox{and } \quad \iota_Y\circ f =f= f\circ \iota_X$$
for every arrow $f:X\mapsto Y$.  
\item
{\em locality}: 
Some of the arrows  are declared      
{\em direct} (or fundamental); they are called {\em links}.
 All arrows $f$ can be 
made from links by composition and adjunction,
 $f=b_n\circ ... \circ b_1, (n\geq 0)$ where $b_i$ are links or
adjoints of links; the empty product $(n=0)$ represents the identity. 

\item
{\em composites:} The objects $X$ are either {\em atomic} or \system s.
In the latter case, $X$ is said to  have internal structure,
and the objects of the \system \ $X$ are called its {\em constituents}. 
\item{\em non-selfinclusion:}
A \system \ cannot be its own object or constituent of an object etc. 
Ultimately, constituents of ... of constituents are atomic. 
\end{enumerate}
A \emph{Semiadditive} \system \  satisfies the same axioms, except that 
arrows may be composed from links and their adjoints with the help of
two operations, $\circ$ and $\oplus$. The $\oplus$-operation 
adds parallel arrows. It makes the 
set $\S (X,Y)$ of all arrows with given source $X$ and target $Y$ 
into an additive (=commutative) semigroup. The distributive law holds
$$ (f_1 \oplus f_2)\circ (g_1 \oplus g_2) = 
f_1\circ g_1 \oplus  f_2\circ g_1 \oplus f_1 \circ g_2 \oplus f_2 \circ g_2$$

An arrow  $o$  is a zero arrow if $f\oplus o = f$ for all $f$.
It is understood that arrows are modulo zero arrows. 
\end{define}
The rationale behind the $\oplus$ operation is that it should be possible 
to interpret two parallel links as a single link, and similarly for arrows. 
\begin{remark}
\label{remark:1}
  In a Semiadditive \system , the set $\S (X,Y)$ 
  can be extended to an additive group if and only if 
  $f\oplus h = g \oplus h $ implies $f=g$, whatever is $h$.
\end{remark}
In the following, we will sometimes  refer to composition with $\circ$ as
{\em multiplication}, and to $\oplus$ as {\em addition} of links.

In addition to the axioms, it is assumed that constituents, 
constituents of constituents etc. of objects of $\S$ are not objects of $\S$.
This assumption is subject to being weakened, but weakening it may 
require adjustment or redesign of software. It would be interesting to weaken 
also the axiomatic property of non-selfinclusion. But this is a
very subtle operation, cp. the discussion in section \ref{sec:splitFork}

We found it convenient to generalize the axiomatic notion of identity arrow 
by introducing {\em identity links} as a special type of  link
which may connect objects that are identical in the sense of indistinguishable.

\begin{define} {\em (Subsystems)}
A {\em subsystem} $\S_1$  of a \system \ $\S$  is generated by a set of 
objects in $\S$ 
and a set of links in $\S$  between these objects.
Its arrows are all arrows in $\S$ that can be composed from these links and
their adjoints. 

The {\em boundary} of $\S_1$ consists of the links in $\S$ with target in 
$\S_1$ which are not links or adjoints of links in $\S_1$. 

The {\em environment} of $\S_1$ is the system generated by the objects of 
$\S$ not in $\S_1$ and the links between them.
\end{define}
In the software, subsystems can be specified by marks in two different ways,
either by marking the (inward) links in its boundary, or by marking the links which belong to it.  

In the description of the software design, we use the following
\begin{nom}$\quad$\\[-4mm]
\label{nom:1}
\begin{description}
\item[Valence] The links with target $X$ are called the \emph{Valences} of $X$
\item[Radical] An object $X$ together with all its valences is called a 
\emph{Radical}.
Given some link $b$, the radical which contains 
its source (object) is called the \emph{source radical} of $b$.   

\item[Membrane] Marks on links which identify the boundary of a subsystem are called
\emph{Membranes}
\item[Path] A \emph{Path} from $X$ to $Y$ is a sequence $b_1,...,b_n$, where 
$b_i$ are links or adjoints of links,
$X$ is the source of $b_1$, $Y$ is the target of $b_n$, and the target
 of $b_i$ is the source of $b_{i+1}$ for $i=1,...,n-1$. The empty
path $(n=0)$ from $X$ to $X$ is identified with $\iota_X$. Paths are special 
subsystems. Marks on links which belong to a certain subsystem are called 
{\bf path-links}.
\end{description}
\end{nom}
In simulations of biological organisms and their parts, membranes can be used as models of cell membranes or envelopes of organs. Path-links are used to model blood vessels and other transport channels, and  neural nets. Attachig enzymes can convert terminal objects of neural nets to sensors and effectors, and objects within blood vessels to pumps. And copy processes can make organisms grow.
Any \system \ with all its interna can be copied by purely local processes, induced e.g. by propagating shocks, cp. section \ref{sec:splitFork}.

\subsection{Basic equations}
In \system s theory, basic equations are often of the form $l=\iota_X$,
where $l:X \mapsto X$ is the arrow specified by some path. 
Maxwell's equations have this form \cite{mack:cmp}.

 Validity of constraints of this form can distinguish between modes of being
 \cite{mack:cmp}. For instance, space time is distiguished by having a
 Lorentzian geometry, 
\footnote{Differential geometry is a father of \system s theory, with parallel transporters as arrows}
 and material bodies are in space and obey conservation laws.

\subsection{Isomorphic systems}
The notion of isomorphic systems is somwhat subtle. Basically, two \system s
$\S^1$ and $\S^2$  are isomorphic if there exists a structure 
preserving map $F:\S^1 \mapsto \S^2 $ which maps 
objects into objects and links into links, and whose inverse exists and has the same property. This induces a map of arrows $f$
into arrows $F(f)$. We require of a structure preserving 
map $F$ that  source and target
of $F(f)$ are the images of source  and target of $f$, and
\footnote{These are the axiomatic properties of functors of a category
\cite{CWM}, with 
the added demand that the map is local in the sense that it maps links to 
links, and preserves the $*$-operation. Anti-isomorphisms are invertible contravariant functors}
\ba
F(f\circ g) &=& F(f) \circ F(g), \label{Functor:1}  \\
F(\iota_X) &=& \iota_{F(X)}  \ . \label{Functor:2} \\
F(f^\ast) &=& F(f)^\ast
\ea
There are also anti-isomorphisms. They relate complementary shapes. For an 
antiisomorphism, eq.(\ref{Functor:1}) is replaced by
\be
F(f\circ g) = F(g) \circ F(f), \label{Functor:3}  \\
\ee
and source and target are interchanged.

It is NOT required that the  internal structure of corresponding objects 
matches when they are not atomic. Nonatomic objects are regarded as 
black boxes. The internal structure of black boxes 
(nonatomic objects) is declared irrelevant when one does not  distinguish
 isomorphic \system s, and so is the distinction between atomic and nonatomic 
 objects. The
only usage of the internal structure is in constructing 
links of the \system \ and their composition $\circ$. One does not look into 
black boxes anymore once they are in place and connected. 

We speak of a {\em strong isomorphism} if it IS demanded that corresponding 
nonatomic objects are strongly isomorphic \system s. (This recursive
definition makes sense because of the axiom of non-selfinclusion.) 
\section{Dynamical systems}
\label{sec:dynamicalSystem}
We consider dynamics in discrete time. 
 A deterministic dynamics of first 
order shall determine a \system \ $\S_{t+1} $ at time $t+1$ and links between 
objects $X$ in $\S_{t+1}$ and objects $Y$ in $\S_t$ from a  given a \system
\ $\S_t$ at time $t$.
If there is a link between $X\in \S_{t+1}$ and $Y\in \S_t$,
$X$ is said to be descendent of $Y$, and $Y$ is ancestor of $X$. 

It is required that 
\begin{enumerate}
\item Every object $X\in \S_{t+1}$ is descendent of at least one object $Y$. 
Conversely, $\{Y_1,...,Y_n\} $ can only be the set of ancestors of some $X$
if the system generated by $Y_1,...,Y_n$ and identity links between them is 
connected. 

If $(n>1)$ we say that there is fusion.
 An object may have more than one descendent ( e.g. copies or translations). 
\item The dynamics is local in the sense that the  isomorphism class
(resp. strong isomorphism class)  
of a subsystem of $\S_{t+1}$ depends only on the \system  \ generated by the 
objects in a 1-neighborhood of its ancestors and the links between them.  
\end{enumerate}
A deterministic dynamics determines $\S_t$ up to isomorphism, a strongly 
deterministic dynamics determines it up to strong isomorphism, i.e. 
including the internal structure of nonatomic objects. 
A dynamics is called universal if it is well defined for every \system \ 
$\S_t$ whatever.

The idea of enzymatic computation is 

\emph{first}, to build dynamics from 
``atomic constituents'' (elementary moves) which we call enzymes.  

\emph{second},
 to incorporate the information about the dynamics into the initial
state by attaching enzymes to links and objects of $\S_t$. 

The enzymes may act conditionally. To this end they are combined with 
predicates which examine the structure of a neigborhood in the system.
A predicate {\tt p} determines a boolean function {\tt p.evaluate(r,v)}
where {\tt r} is a radical,
and  {\tt v} is (pointer to) a valence, typically of {\tt r}.
The pair (predicate, enzyme) is also called a {\em mechanism}. The enzymes 
are abstractions not only of the biochemical enzymes \cite{Alberts} which 
govern life processes in cells, but of any agent of change. 
Biochemical enzymes show specificity \cite{Alberts}, i.e. they come with their
 predicates inseparably build in.
 We shall occasionally speak of enzymes when we really mean
mechanisms. 

There are several kinds of basic enzymes, i.e. elementary moves. 
\begin{description}
\item [motion] An arrow becomes promoted to link.  Either

a) The composite of two (or more) links becomes a link, i.e. 
indirect relations become direct. \emph{example 1: Friend of a friend becomes 
friend. example 2: Equations of motion of fundamental physics like
 Maxwells  equations \cite{mack:cmp,mack:cargese} or motion of particles in space.[Transport of material involves motion of particles in space (or an effective decription of that in terms of algebraic values)].
 example 3: catalysis, figure 
\ref{fig:catalysis}. example 4: logical deductions, cp. 
ref.\cite{mack:cmp,schrattenholzer}}.

\begin{figure}[b!]
  \begin{center}
    \leavevmode
     \input{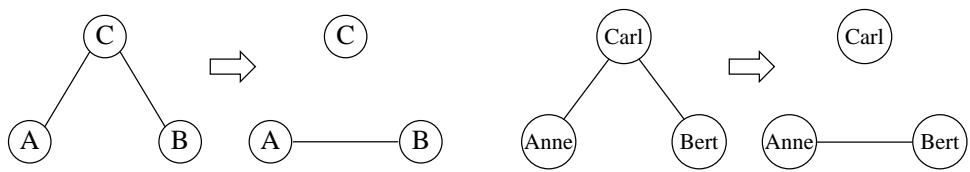}
  \end{center}
  \caption{\small Catalysis in chemistry and elsewhere. A catalyst {\tt C} binds molecules {\tt A} and {\tt B}.
First a substrate-enzyme complex is built, where {\tt A} and {\tt B} bind to {\tt C}. Next the composite arrow from
{\tt A} to {\tt B} becomes fundamental.}
  \label{fig:catalysis}
\end{figure}

b) The adjoint of a link which was not a link becomes a link.
 \emph{example: reciprocation}

The converse - adjoint of a link loses the status  of a link - is also subsumed
under motion. Motion is reversible.
\item[growth] An object has two or more descendents - e.g copies which are made
or taken from the environment.  Fusion of objects is also subsumed under growth, and so is the making of an object with internal structure from a subsystem.
Growth is reversible if  descendents are connected by identity links 
(cp. section \ref{sec:system} after definition \ref{def:system}).
Removal of the identity link would  be subsumed under death, s. below.
\item [death] An object may disappear with its links, links may disappear 
 together with  their adjoints. Death is usually irreversible.

\item[cognition] A new link is made between objects $X_1$, $X_2$ with matching internal structure. Locality requires that $X_1$, $X_2$ are connected by 
a path of short length in $\S_t$, i.e. there is a preexisting relation
between them. But the 
new link need not be the arrow determined by the preexisting path.
\emph{example: a chemical bond may be established when there was
spacial proximity before.}  Links made this way are called 
{\em cognitive links}.
\item[enzyme management] Enzymes are attached to  objects and links of 
$\S_{t+1}$, or their action somewhere at time $t+1$ is specified in other ways.
\end{description}  
The making of cognitive links only makes sense if the dynamics determines
future \system s up to {\em strong} isomorphism, because the internal 
structure of nonatomic objects matters. Cognitive links can be of several 
types $f$. $f$ is a boolean function (``matching structure'') of two objects,
 typically determined
by a predicate $p$ with the property that $p(r,v)$ depends only on 
the object of radical $r$ and the source object of valence $v$. 
Its return value is also supposed to be obtained by enzymatic computation
(with access to the internal structure of objects which are \system s). 

The most important example of matching structure are isomorphisms and
antiisomorphism. The latter implement the {\em lock key} mechanism
which is responsible for specificity in biochemistry \cite{Alberts}.  
Determining isomorphisms of graphs is an NP-hard problem. But this is not very 
relevant here because in practise one will encounter graphs with special 
properties. Matching structures can also be detected by neural nets, here
implemented as subsystems made with path-links.

We will subsume the making of identity links 
between indistinguishable atomic objects under cognition as well, and
regard such identity links as prototypical examples of cognitive links.

A description of the most important enzymes will be given later.

So far we talked about deterministic dynamics. In a stochastic dynamics, 
enzymes act with certain probabilities. 
\subsection{Concurrency}
\label{sec:Concurrency}
The action of enzymes at different locations will not commute in general. 
This results in what is known to computer scientists working on parallel 
computing as the {\em concurrency problem}.
 Petri nets are a well know example \cite{Petri}.
In the spirit of biology, one may just ignore this problem, admitting 
some randomness or indeterminacy in the dynamics, e.g. through a 
hidden  dependence
on an ordering of the valences of a radical.  But it is useful to have the 
option of specifying a well defined deterministic dynamics for \system s. 

We propose to  achieve this by a generalization of
 Jacobi sweeps.

Let us first recall the notion of a Jacobi sweep. Consider a grid made 
of nodes (objects) connected by links, where the nodes and/or the links 
carry some data. Given a cost function whose arguments are the 
aforementioned data and which is a sum of contributions which depend only
on a neighborhood of individual nodes or links. One may attempt its 
minimization by relaxation. One makes sweeps through the grid, updating
data at individual nodes or links such that the cost function is minimized
under the constraint that all other data remain frozen. In a Jacobi sweep
(as opposed to Gauss Seidel sweeps)\cite{MGrid},
the data after sweep $t+1$ are determined solely by the data after sweep $t$. 
As a result it does not matter in which order the nodes and links of the grid 
are visited - the effect of the individual updating operations will commute.

We may proceed in the same way, except that we divide the making of
the \system \ $\S_{t+1} $ at time $t$ from the \system \ $\S_t$ at time $t$
into several steps which take place at times $t+\frac 1 4$, 
$t+\frac 1 2$, $t+\frac 3 4$ and $t+1$. 
In intermediate steps, there will be links connecting objects 
in $\S_t$ with objects in $S_t+1$. We call them ``time links''

We may divide enzymes into classes according to whether they act at 
time $t+\frac 14$, $t+\frac 12$ or $t+1$. The intermediate step at 
time $t+\frac 34 $ was introduced for pedagogical reasons only. The 
enzymes which act at time $t+\frac 1 4 $ will be called ``object-making'', 
those at time $t+\frac 12 $ ``link-making''.
 The link-making enzymes may also put marks on 
newly made links to indicate information on their adjoints.

As explained in section \ref{sec:dynamicalSystem}, 
our processes are such that every object $Y$ in $\S_{t+1}$ is    
descendent of (at least) one object $X$ in $\S_t$. 

\begin{figure}
\begin{center}
\setlength{\unitlength}{0.6mm}
\begin{picture}(120,120)(-10,0)


\put(90,120){$t$}

\put(20,100){\begin{picture}(0,0)
\put(20,20){\circle*{2}}
\put(-5,14){\circle*{2}}        
\put(5,25){\circle*{2}}        
\put(18,19.5){\vector(-4,-1){21}}
\put(5,25){\vector(3,-1){13}}
\end{picture}}

\put(40,120){\begin{picture}(0,0)
\put(0,0){\circle*{2}}
\put(20,0){\circle*{2}}
\put(2,-0.5){\vector(1,0){16}}
\put(18,0.5){\vector(-1,0){16}}
\end{picture}}

\put(60,120){\begin{picture}(0,0)
\put(0,0){\circle*{2}}
\put(20,0){\circle*{2}}
\put(2,-0.5){\vector(1,0){16}}
\put(18,0.5){\vector(-1,0){16}}
\end{picture}}

\put(-5,114){\begin{picture}(0,0)
\put(0,0){\circle*{2}}
\put(20,0){\circle*{2}}
\put(2,-0.5){\vector(1,0){16}}
\put(18,0.5){\vector(-1,0){16}}
\end{picture}}

\put(5,125){\begin{picture}(0,0)
\put(0,0){\circle*{2}}
\put(20,0){\circle*{2}}
\put(2,-0.5){\vector(1,0){16}}
\put(18,0.5){\vector(-1,0){16}}
\end{picture}}  

\put(90,85){$t+\frac 14$}

\put(20,60){\begin{picture}(0,0)
\put(20,20){\circle*{2}}
\put(-5,14){\circle*{2}}        
\put(5,25){\circle*{2}}        
\put(18,19.5){\vector(-4,-1){21}}
\put(5,25){\vector(3,-1){13}}
\end{picture}}
 
\put(40,80){\begin{picture}(0,0)
\put(0,0){\circle*{2}}
\put(20,0){\circle*{2}}
\put(2,-0.5){\vector(1,0){16}}
\put(18,0.5){\vector(-1,0){16}}
\end{picture}}

\put(60,80){\begin{picture}(0,0)
\put(0,0){\circle*{2}}
\put(20,0){\circle*{2}}
\put(2,-0.5){\vector(1,0){16}}
\put(18,0.5){\vector(-1,0){16}}
\end{picture}}

\put(-5,74){\begin{picture}(0,0)
\put(0,0){\circle*{2}}
\put(20,0){\circle*{2}}
\put(2,-0.5){\vector(1,0){16}}
\put(18,0.5){\vector(-1,0){16}}
\end{picture}}

\put(5,85){\begin{picture}(0,0)
\put(0,0){\circle*{2}}
\put(20,0){\circle*{2}}
\put(2,-0.5){\vector(1,0){16}}
\put(18,0.5){\vector(-1,0){16}}
\end{picture}}
\put(40,80){\begin{picture}(0,0)
\put(0,0){\circle*{2}}
\put(-5,14){\circle*{2}}        
\put(5,25){\circle*{2}}        
\put(5.5,22){\vector(-1,-4){5}}
\put(-1,2){\vector(-1,3){3.3}}
\end{picture}}

\put(60,80){\begin{picture}(0,0)
\put(0,0){\circle*{2}}
\put(0,20){\circle*{2}}
\put(-0.5,2){\vector(0,1){16}}
\put(0.5,18){\vector(0,-1){16}}
\end{picture}}

\put(80,80){\begin{picture}(0,0)
\put(0,0){\circle*{2}}
\put(0,20){\circle*{2}}
\put(-0.5,2){\vector(0,1){16}}
\put(0.5,18){\vector(0,-1){16}}
\end{picture}}

\put(25,85){\begin{picture}(0,0)
\put(0,0){\circle*{2}}
\put(0,20){\circle*{2}}
\put(-0.5,2){\vector(0,1){16}}
\put(0.5,18){\vector(0,-1){16}}
\end{picture}}

\put(15,74){\begin{picture}(0,0)
\put(0,0){\circle*{2}}
\put(0,20){\circle*{2}}
\put(-0.5,2){\vector(0,1){16}}
\put(0.5,18){\vector(0,-1){16}}
\end{picture}}

\put(5,85){\begin{picture}(0,0)
\put(0,0){\circle*{2}}
\put(0,20){\circle*{2}}
\put(-0.5,2){\vector(0,1){16}}
\put(0.5,18){\vector(0,-1){16}}
\end{picture}}

\put(-5,74){\begin{picture}(0,0)
\put(0,0){\circle*{2}}
\put(0,20){\circle*{2}}
\put(-0.5,2){\vector(0,1){16}}
\put(0.5,18){\vector(0,-1){16}}
\end{picture}}


\put(90,50){$t+\frac 12$}
\put(20,20){\begin{picture}(0,0)
\put(20,20){\circle*{2}}
\put(-5,14){\circle*{2}}        
\put(5,25){\circle*{2}}        
\put(18,19.5){\vector(-4,-1){21}}
\put(5,25){\vector(3,-1){13}}
\end{picture}}

\put(40,40){\begin{picture}(0,0)
\put(0,0){\circle*{2}}
\put(20,0){\circle*{2}}
\put(2,-0.5){\vector(1,0){16}}
\put(18,0.5){\vector(-1,0){16}}
\end{picture}}

\put(60,40){\begin{picture}(0,0)
\put(0,0){\circle*{2}}
\put(20,0){\circle*{2}}
\put(2,-0.5){\vector(1,0){16}}
\put(18,0.5){\vector(-1,0){16}}
\end{picture}}

\put(-5,34){\begin{picture}(0,0)
\put(0,0){\circle*{2}}
\put(20,0){\circle*{2}}
\put(2,-0.5){\vector(1,0){16}}
\put(18,0.5){\vector(-1,0){16}}
\end{picture}}

\put(5,45){\begin{picture}(0,0)
\put(0,0){\circle*{2}}
\put(20,0){\circle*{2}}
\put(2,-0.5){\vector(1,0){16}}
\put(18,0.5){\vector(-1,0){16}}
\end{picture}}
\put(40,40){\begin{picture}(0,0)
\put(0,0){\circle*{2}}
\put(-5,14){\circle*{2}}        
\put(5,25){\circle*{2}}        
\put(5.5,22){\vector(-1,-4){5}}
\put(-1,2){\vector(-1,3){3.3}}
\put(-1,2){\vector(-1,3){3.3}}\end{picture}}

\put(60,40){\begin{picture}(0,0)
\put(0,0){\circle*{2}}
\put(0,20){\circle*{2}}
\put(-0.5,2){\vector(0,1){16}}
\put(0.5,18){\vector(0,-1){16}}
\end{picture}}

\put(80,40){\begin{picture}(0,0)
\put(0,0){\circle*{2}}
\put(0,20){\circle*{2}}
\put(-0.5,2){\vector(0,1){16}}
\put(0.5,18){\vector(0,-1){16}}
\end{picture}}

\put(25,45){\begin{picture}(0,0)
\put(0,0){\circle*{2}}
\put(0,20){\circle*{2}}
\put(-0.5,2){\vector(0,1){16}}
\put(0.5,18){\vector(0,-1){16}}
\end{picture}}

\put(15,34){\begin{picture}(0,0)
\put(0,0){\circle*{2}}
\put(0,20){\circle*{2}}
\put(-0.5,2){\vector(0,1){16}}
\put(0.5,18){\vector(0,-1){16}}
\end{picture}}

\put(5,45){\begin{picture}(0,0)
\put(0,0){\circle*{2}}
\put(0,20){\circle*{2}}
\put(-0.5,2){\vector(0,1){16}}
\put(0.5,18){\vector(0,-1){16}}
\end{picture}}

\put(-5,34){\begin{picture}(0,0)
\put(0,0){\circle*{2}}
\put(0,20){\circle*{2}}
\put(-0.5,2){\vector(0,1){16}}
\put(0.5,18){\vector(0,-1){16}}
\end{picture}}

\put(60,40){\begin{picture}(0,0)
\put(2,2){\vector(1,1){16}}
\end{picture}}

\put(80,40){\begin{picture}(0,0)
\put(-2,2){\vector(-1,1){16}}
\end{picture}}

\put(5,45){\begin{picture}(0,0)
\put(2,2){\vector(1,1){16}}
\end{picture}}

\put(25,45){\begin{picture}(0,0)
\put(-2,2){\vector(-1,1){16}}
\end{picture}}

\put(40,40){\begin{picture}(0,0)
\put(2,2){\vector(1,1){16}}
\end{picture}}

\put(60,40){\begin{picture}(0,0)
\put(-2,1){\vector(-3,2){20.5}}
\end{picture}}

\put(15,34){\begin{picture}(0,0)
\put(-2,2){\vector(-1,1){16}}
\end{picture}}

\put(-5,34){\begin{picture}(0,0)
\put(2,2){\vector(1,1){16}}
\end{picture}}

\put(35,54){\begin{picture}(0,0)
\put(-18,-18){\vector(1,1){16}} 
\end{picture}}

\put(40,40){\begin{picture}(0,0)
\put(-2,1.2){\vector(-1,2){11}}
\end{picture}}



\put(90,20){$t+1$}
\put(20,-20){\begin{picture}(0,0)
\put(20,20){\circle*{2}}
\put(-5,14){\circle*{2}}        
\put(5,25){\circle*{2}}        
\put(18,19.5){\vector(-4,-1){21}}
\put(5,25){\vector(3,-1){13}}
\end{picture}}

\put(40,0){\begin{picture}(0,0)
\put(0,0){\circle*{2}}
\put(20,0){\circle*{2}}
\put(2,-0.5){\vector(1,0){16}}
\put(18,0.5){\vector(-1,0){16}}
\end{picture}}

\put(60,0){\begin{picture}(0,0)
\put(0,0){\circle*{2}}
\put(20,0){\circle*{2}}
\put(2,-0.5){\vector(1,0){16}}
\put(18,0.5){\vector(-1,0){16}}
\end{picture}}

\put(-5,-6){\begin{picture}(0,0)
\put(0,0){\circle*{2}}
\put(20,0){\circle*{2}}
\put(2,-0.5){\vector(1,0){16}}
\put(18,0.5){\vector(-1,0){16}}
\end{picture}}

\put(5,5){\begin{picture}(0,0)
\put(0,0){\circle*{2}}
\put(20,0){\circle*{2}}
\put(2,-0.5){\vector(1,0){16}}
\put(18,0.5){\vector(-1,0){16}}
\end{picture}}
\put(40,0){\begin{picture}(0,0)
\put(0,0){\circle*{2}}
\put(-5,14){\circle*{2}}        
\put(5,25){\circle*{2}}        
\put(5.5,22){\vector(-1,-4){5}}
\put(-1,2){\vector(-1,3){3.3}}
\end{picture}}

\put(60,0){\begin{picture}(0,0)
\put(0,0){\circle*{2}}
\put(0,20){\circle*{2}}
\put(-0.5,2){\vector(0,1){16}}
\put(0.5,18){\vector(0,-1){16}}
\end{picture}}

\put(80,0){\begin{picture}(0,0)
\put(0,0){\circle*{2}}
\put(0,20){\circle*{2}}
\put(-0.5,2){\vector(0,1){16}}
\put(0.5,18){\vector(0,-1){16}}
\end{picture}}

\put(25,5){\begin{picture}(0,0)
\put(0,0){\circle*{2}}
\put(0,20){\circle*{2}}
\put(-0.5,2){\vector(0,1){16}}
\put(0.5,18){\vector(0,-1){16}}
\end{picture}}

\put(15,-6){\begin{picture}(0,0)
\put(0,0){\circle*{2}}
\put(0,20){\circle*{2}}
\put(-0.5,2){\vector(0,1){16}}
\put(0.5,18){\vector(0,-1){16}}
\end{picture}}

\put(5,5){\begin{picture}(0,0)
\put(0,0){\circle*{2}}
\put(0,20){\circle*{2}}
\put(-0.5,2){\vector(0,1){16}}
\put(0.5,18){\vector(0,-1){16}}
\end{picture}}

\put(-5,-6){\begin{picture}(0,0)
\put(0,0){\circle*{2}}
\put(0,20){\circle*{2}}
\put(-0.5,2){\vector(0,1){16}}
\put(0.5,18){\vector(0,-1){16}}
\end{picture}}

\put(40,0){\begin{picture}(0,0)
\put(20,20){\circle*{2}}
\put(-5,14){\circle*{2}}        
\put(5,25){\circle*{2}}        
\put(18,19.5){\vector(-4,-1){21}}
\put(5,25){\vector(3,-1){13}}
\end{picture}}

\put(60,20){\begin{picture}(0,0)
\put(0,0){\circle*{2}}
\put(20,0){\circle*{2}}
\put(2,-0.5){\vector(1,0){16}}
\put(18,0.5){\vector(-1,0){16}}
\end{picture}}

\put(-5,14){\begin{picture}(0,0)
\put(0,0){\circle*{2}}
\put(20,0){\circle*{2}}
\put(2,-0.5){\vector(1,0){16}}
\put(18,0.5){\vector(-1,0){16}}
\end{picture}}

\put(15,14){\begin{picture}(0,0)
\put(0,0){\circle*{2}}
\put(20,0){\circle*{2}}
\put(2,-0.5){\vector(1,0){16}}
\put(18,0.5){\vector(-1,0){16}}
\end{picture}}

\put(5,25){\begin{picture}(0,0)
\put(0,0){\circle*{2}}
\put(20,0){\circle*{2}}
\put(2,-0.5){\vector(1,0){16}}
\put(18,0.5){\vector(-1,0){16}}
\end{picture}}

\put(25,25){\begin{picture}(0,0)
\put(0,0){\circle*{2}}
\put(20,0){\circle*{2}}
\put(2,-0.5){\vector(1,0){16}}
\put(18,0.5){\vector(-1,0){16}}
\end{picture}}


\end{picture} 
\end{center}
\caption{splitfork dynamics, concurrent version}
\label{fig:concur}
\end{figure}
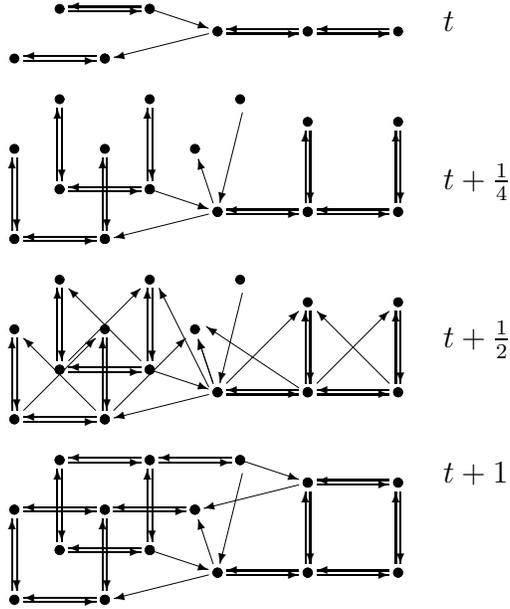

Ignoring the attachment of enzymes to $\S_{t+1}$ at first, 
the course of events is like this (cp. figure \ref{fig:concur} 
for an example) .
\begin{description}
\item[$t+\frac 14$] Make descendents $Y_1,...,Y_n \in \S_t$ of \emph{objects}
 $X\in \S_t$, if any; $(n\leq 2)$. 
Connect them by time links. When $(n>1)$ make identity links
between identical copies $Y_1,...,Y_n$ if demanded by an enzyme.
\item[$t+\frac 12$] Make \emph{links} that will end up in $\S_{t+1}$. 
Every link-making enzyme may contribute some link or links, in a manner which 
depends on its neighborhood in $\S_t$. For pedagogical reasons, the links are 
 made at time $t+\frac 12 $ to connect one object in $\S_t$ and one object 
in $\S_{t+1}$, and at time 
\item[$t+\frac 34 $] the ends in $\S_t$ of such links are lifted to 
$\S_{t+1}$ by composition with a time link. 
\item[$t+1$] 
Depending on the marks on the newly made links in $\S_{t+1}$, adjointness 
relations are established among links, or new fundamental adjoints are made.  
The operation is local in that it involves only the individual pairs of 
sets $\S(Y,Y^\prime)$ and $\S(Y^\prime, Y)$ of links between given objects
$Y, Y^\prime $. 
\end{description}

So far we neglected to say how the enzymes are put  into the \system \ $\S_{t+1}$.
Object making enzymes are attached to descendents or identity links between
them in step $t+\frac 14$.
Link-making enzymes are attached to links in step $t+\frac 12$.

Dynamics of this kind is well defined because all operations within anyone 
of the steps commute. 

In the end, the time links may be removed (if desired).
This step is not shown in figure 
\ref{fig:concur}.
 
Unfortunately the use of Jacobi sweeps is less convenient than 
admitting a hidden dependence 
of the order in which enzymes at neighbouring locations act on an ordering of
 valences in  radicals.
This is so because the composition of chains of enzymes which act one after 
another is not a simple matter anymore.  

\section{The System Class Laboratory}
The System Class Laboratory (SyCL) is a software package composed of three 
parts
\begin{description}
\item[core] package is a $C^{++}$ class library which encodes the basic 
functionality demanded by the axioms of a \emph{semi-additive system}
and enzymatic dynamics on it. 
\item[presentation] package includes a graphic interface. Using it 
one can construct and 
manipulate such \system s by mouse-click, invoking the conditional action of
basic enzymes and of compound enzymes made from them. 
The presentation package also includes to translate \system s to and from
{\bf XML}, i.e. text. 
\item[interpreter] and parser package is based on a LISP-like scripting 
language. One can write programs in this language and run them. 
Alternatively, operations carried out by mouse click on the graphic interface
are automatically recorded as commands of the scripting language. 
In this way one may compose programs by mouse click.      
\end{description}

\section{The class library}
\subsection{Basic Data Types}
\label{sec:basicData}
In principle, \system \ theory is self-contained 
\footnote{in contrast with automata theory \cite{automata}}.
There are no data
in \system s other than their structure, and no states of any part of a 
{\sc System} other than
its structure. The miracle is how much can be modeled with so little building 
material.

In practice, it is nevertheless convenient to admit data inside objects and 
links which can be thought to code for internal structure in some way. 
(Links with internal structure can be thought of as \system 's with two 
interfaces)

The data inside links are instances of a class \emph{AlgebraicValue}. 
They admit algebraic operations. They can be 
added, multiplied and multiplied with real numbers in a manner which is
described below in subsection \ref{sec:algebraicValue}.
Their equality can also be ascertained.  The class
\emph{AlgebraicValue} is derived from a class \emph{Value} which does
not have the algebraic operations as methods. Objects can contain arbitrary 
\emph{Value}s.  Multiplication and addition of algebraic values
that reside inside links is invoked when links are composed using 
the axiomatic $\circ$ and $\oplus$-operations. 
Several different classes are derived from AlgebraicValue, including 
the class Predicate. We mention it separately because of its important role.  
Enzymes may also be attached to objects and links in the guise of
EnzymaticValue, also derived from AlgebraicValue.

The other data types correspond with Nomenclature \ref{nom:1}, except for the
 following implementational detail. The adjoint of a link $b$ 
 with source $X$ which is not itself a link is nevertheless included in the 
list of Valences of the radical containing $X$, and is marked as a 
``virtual valence''.  This is 
technically convenient because the target of $b$ can be addressed as source of
 its adjoint. 

The basic data types are 
\begin{description}
\item[Value, AlgebraicValue] contain text or numerical data with operations as
 described

\item[Objects] may contain a singly linked list of \emph{Values}.
 They can be copied.
\item[Valences] contain a pointer to their \emph{source radical},
 a pointer to their  \emph{adjoint}s,
 and possibly: a singly linked list of \emph{AlgebraicValue}'s, a 
 singly linked list of \emph{Membrane}s, a singly linked list of \emph{Path}s.

 By definition, the adjoint  is a valence of the source radical,
 possibly virtual. 

 Composition $\circ$ of a valence with valences of its source radical is 
defined. 

 Parallel valences may be added to a single valence. 
\item[Radicals] contain an object, a doubly linked list of valences, and possibly  a position in 3-dimensional space.   
\item[Predicates] determine a boolean function which takes as argument a pair
 {\tt(r,v)},
  where {\tt v} is a pointer to valence (maybe NULL) and {\tt r} is a radical
(typically the target of {\tt v}).
Predicates can be composed with junctors ``and, or, not''. Among the 
predicates are keys which permit to pass membranes selectively.
\item[Membranes] are attached to valences.  They mark the boundary 
of a subsystem. They carry a string or a {\tt void*} pointer as code to
 distinguish them. 
\item[PathLinks] contain a reference to a valence and data to identify the
 precursor, successor and adjoint pathLink.
 A valence knows the pathLinks that
 pass through it, cp. above. 
\item[System] A connected nonempty \system \ is identified by a pointer
 {\tt root} to 
one of its radicals, and possibly by the code of a membrane which bounds it. 

From {\tt root} one can access the source radicals of its valences, and so on. 
All the information on the \system \ is in the radicals which can be accessed 
in this way. 
\item[Enzyme] Enzymes have a method {\tt Radical \&  act(r,v)} which takes as 
  its arguments  a radical 
{\tt r} and a pointer {\tt v} to a pointer *v  to  a valence (of {\tt r}).
{\tt *v} may be NULL. The action is  on a neighborhood of {\tt r}. {\tt *v} may be changed to point to a different valence, therefore a pair 
{\tt (Radical \&, Valence *)} is effectively returned, which can be used for input to another enzyme. 
\end{description}
One may define a {\em path distance} $d$ between objects $X$ and $Y$ by use 
of  paths $\{ b_1, ... , b_n\} $. $d$ is the minimal value of
 $n$ among all the  paths from $X$ to $Y$. 

\subsection{AlgebraicValues}
\label{sec:algebraicValue}
The implementation of algebraic operations with AlgebraicValues makes essential
use of object oriented design features like inheritance and virtuality. 
\cite{GHJV}. 
We assume that the reader is familiar with these. 

The algebraic operations satisfy the usual laws of associativity and
distributivity. Multiplication may be noncommutative, and in one 
special case
(the max-plus-''algebra''), subtraction is undefined. 

There is an inheritance tree which specifies how different classes
of algebraic values are derived from the base class \emph{AlgebraicValue}. 
Each class has its (virtual) methods for multiplication with members 
of the same class from the right or left, for addition, and for multiplication 
with real numbers $\alpha$. In some cases (like strings) 
the multiplication with  real
$\alpha $ does nothing if $\alpha \neq 0$, and converts to a 
zero-element otherwise. Given any two classes $A$, $B$ of algebraic values,
there is a unique class $C$, the `` least common ancestor``,
 such that 
$A$ and $B$ are derived from $C$, but not from any class that is derived from 
$C$. The result of right multiplication $*=$ is
computed by invoking the multiplication-method of $C$. Addition and left multiplication are treated in the same way.  

\begin{figure}[h!]
  \begin{center}
    \leavevmode
    \input{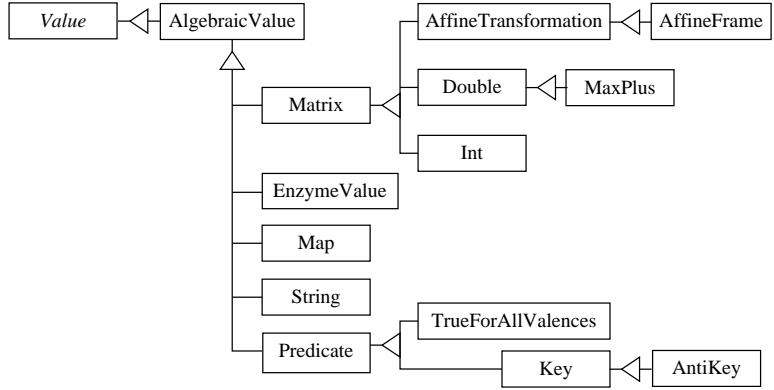}
  \end{center}
  \caption{\small The class diagram for different types of values.}
  \label{basicclasses}
\end{figure}

The inheritance tree is shown in figure \ref{basicclasses}. 

There is one type of algebraic values of interest where the condition
of remark \ref{remark:1} is 
violated, the {\em max-plus} or {\em timetable} ``algebra''. Its data 
are real numbers (or real matrices), $a\circ b = a + b$ (addition of real
numbers) and $a \oplus b = max(a,b)$. For matrices, the operations are 
entry-wise.      
This ``algebra'' is useful in so called \emph{discrete event systems} 
\cite{DED},
and in optimization problems where one requires the maximum of errors 
which are given by real return values of local cost functions.

There may be  a (singly linked) list of AlgebraicValues in a link. 
In this case, multiplication is element by element in the list.  

Values in Objects need not be of type {\em AlgebraicValue}. But for 
instances $\eta$ of useful data types, multiplication with 
AlgebraicValue $A$ from the left is defined
and is associative (i.e. $(A * B)*\eta = A*(B*\eta))$. The multiplication may be 
trivial, i.e. $A * \eta = \eta $ for all $A$.  In case $\eta $ is itself of 
type AlgebraicValue, multiplication is defined as described above. 

\begin{define}(\emph{Parallel Transport})
Given a link $b$ with AlgebraicValue $A$ and the value $\eta $ of its source
object $Y$, $A*\eta$ is regarded as a potential value of the 
target object $X$ of the link $b$, and the operation is called the 
{\em parallel transport} of $\eta$ from $Y$ to $X$ {\em along the link}
 $b$. 
\end{define}

The parallel 
transport along links may be composed to parallel transport along paths 
$\{b_0,..., b_n\}$. Parallel transport along $\iota_Y$ is the trivial
map $\eta\mapsto \eta$. The notion comes from lattice gauge theory
of elementary  particles \cite{Creutz} and general relativity. 

\begin{remark}
Because of associativity, the result of the parallel 
transport depends actually only on the arrow defined by the path.  
\end{remark}

It may happen that the result of
parallel transport of some type of value from $Y$ to $X$ is independent of the
 chosen path from $Y$ to $X$. In this case the type of value is called an
 \emph{invariant}.  

Invariants have a global meaning
in a connected \system , because they may be parallel transported 
from any $Y$ to any $X$ in a unique way. Important examples are 
i) quantities of any chemical substance which is transferred 
from object to object by diffusion, ii) payments in an economy. 

Counterexamples are any kind of ``non-ideal communication'', such as 
communication of humans in natural language. Utterances in natural language 
do not generally have a globally well defined meaning - one's owl is 
the other's nightingale. Gossip is an example. When the message comes 
back to speaker $X$ who send it out, it may have undergone dramatic change. So
the result of communication around some loop
is not the same as along the identity $\iota_X$.  

\subsection{Enzymes}
\label{sec:enzymes}
We proceed to describe the most important micro-enzymes, beginning with 

{\em motion}: It involves either multiplication of two links, or promotion of a virtual adjoint of a link to the status of a link. 
We have enzymes $\_ VML$ for left multiplication and 
$\_ AMR$ for right multiplication. Given a $\_ VML$-enzyme {\tt e} with
attached predicate {\tt p}, {\tt e.act(r,m)} will go through all valences
{\tt v} of {\tt r}, and multiplies those {\tt v} from the left with the
 adjoint 
of {\tt m} for which {\tt p.evaluate(r,v)} returns TRUE. An illustration is 
found below. In this figure, virtual valences are indicated as 
dotted lines. The $\_ AMR$-enzyme is similar. 

\noindent
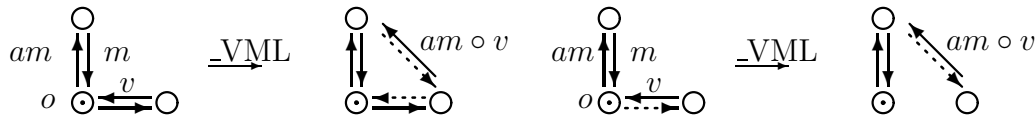
\begin{center}
\setlength{\unitlength}{0.7mm}
\begin{picture}(190,24)
\thicklines
\put(8,1) {\makebox{$o$}}
\put(2,10) {\makebox{$am$}}
\put(20,10) {\makebox{$m$}}
\put(23,4) {\makebox{$v$}}
\put(16,2) {\circle{4}} \put(16,2) {\circle{0.5}}
\put(32,2) {\circle{4}}
\put(16,18) {\circle{4}}
\put(15,5) {\vector(0,1){10}}
\put(17,15) {\vector(0,-1){10}}
\put(29,3) {\vector(-1,0){10}}
\put(19,1) {\vector(1,0){10}}
\thinlines
\put(40,10) {\makebox{\_VML}}
  \put(40,9) {\vector(1,0){10}}
\thicklines
\put(80,12) {\makebox{$am \circ v$}}
\put(68,2) {\circle{4}} \put(68,2) {\circle{0.5}}
\put(84,2) {\circle{4}}
\put(68,18) {\circle{4}}
\put(67,5) {\vector(0,1){10}}
\put(69,15) {\vector(0,-1){10}}
\put(73,3) {\vector(-1,0){2}} \dottedline{2}(73,3)(81,3)
\put(71,1) {\vector(1,0){10}}
\put(83,7) {\vector(-1,1){10}}
\put(81,7) {\vector(1,-1){2}} \dottedline{2}(73,15)(81,7)
\thicklines
\put(110,1) {\makebox{$o$}}
\put(105,10) {\makebox{$am$}}
\put(120,10) {\makebox{$m$}}
\put(123,4) {\makebox{$v$}}
\put(116,2) {\circle{4}} \put(116,2) {\circle{0.5}}
\put(132,2) {\circle{4}}
\put(116,18) {\circle{4}}
\put(115,5) {\vector(0,1){10}}
\put(117,15) {\vector(0,-1){10}}
\put(129,3) {\vector(-1,0){10}}
\put(127,1) {\vector(1,0){2}} \dottedline{2}(119,1)(127,1)
\thinlines
\put(140,10) {\makebox{\_VML}}
  \put(140,9) {\vector(1,0){10}}
\thicklines
\put(180,12) {\makebox{$am \circ v$}}
\put(168,2) {\circle{4}} \put(168,2) {\circle{0.5}}
\put(184,2) {\circle{4}}
\put(168,18) {\circle{4}}
\put(167,5) {\vector(0,1){10}}
\put(169,15) {\vector(0,-1){10}}
\put(183,7) {\vector(-1,1){10}}
\put(181,7) {\vector(1,-1){2}} \dottedline{2}(173,15)(181,7)
\end{picture}
\end{center}


Alternatively, motion may involve the promotion of the adjoint of a link 
to the status of link. There are enzymes to do that, one of them is the
$\_ MAD$-enzyme. It makes adjoints of all valences into links
which fulfill the condition set by a predicate. $\_ RFU $ acts in the same 
way, except that the argument {\tt v} is replaced by its adjoint, and 
the source of {\tt v} substitutes for {\tt r}.

 \begin{center}
\setlength{\unitlength}{0.8mm}
\begin{picture}(130,10)
\thicklines
\put(16,2) {\circle{4}} \put(16,2) {\circle{0.5}}
\put(5,1) {\vector(-1,0){2}} \dottedline{2}(5,1)(13,1)
\put(3,3) {\vector(1,0){10}}
\put(21,1) {\vector(-1,0){2}} \dottedline{2}(21,1)(29,1)
\put(19,3) {\vector(1,0){10}}
\thinlines
\put(34,1) {\makebox{\_MAD}}
  \put(34,0) {\vector(1,0){10}}
\thicklines
\put(68,2) {\circle{4}} \put(68,2) {\circle{0.5}}
\put(65,1) {\vector(-1,0){10}}
\put(55,3) {\vector(1,0){10}}
\put(73,1) {\vector(-1,0){2}} \dottedline{2}(73,1)(81,1)
\put(71,3) {\vector(1,0){10}}
\thinlines
\put(86,1) {\makebox{\_RFU}}
  \put(86,0) {\vector(1,0){10}}
\thicklines
\put(118,2) {\circle{4}} \put(118,2) {\circle{0.5}}
\put(115,1) {\vector(-1,0){10}}
\put(105,3) {\vector(1,0){10}}
\put(131,1) {\vector(-1,0){10}}
\put(121,3) {\vector(1,0){10}}
\end{picture}
\end{center}

To change a valences status from fundamental to virtual, an enzyme $\_MVV$ is used.

 {\em Growth}: The most important enzyme is $\_ CPO$ which copies an object,
and links original and copy by an identity link. 
(Another version searches for the copy in the environment). The object may be composite, i.e. a \system . 
$\_ CPO $ appears in the first micro-enzyme in the splitFork enzyme as
shown in figure \ref{fig:microFork} in section \ref{sec:splitFork}. 
In the figure there, some predicates are indicated by a string-code
$\{ ? ... \}$ enclosed in braces.  Their meaning is as follows

$?nLK$ {\tt v} is not a link (i.e. is a virtual valence)

$?iAD$ adjoint of {\tt v} is a link, negation $?nAD$.

$?iID$ {\tt v} is an identity link.

$?hLS$ detect a triangular structure.

There are also enzymes like $\_SYA $ which make objects of subsystems whose boundary is marked by membranes.

{\em Death}:  links are removed by $\_ RTV$ or $\_ RMV$ enzymes which kill the
 particular valence given as argument, or all valences which fulfill a 
condition. (The $\_ RMV$-enzyme operates in figure \ref{fig:microFork}
 to remove identity links.) 

A removal of a radical with all its valences and their adjoints by 
the $\_ DEL$-enzyme.

{\em Enzyme management} enzymes: Among them 
 there are presentation enzymes.
They have  relatives in  cell biology \cite{Alberts}.   They hand
arguments (r,v) to another enzyme. For instance, for a $\_ PRS$-enzyme
 {\tt e}, (``present Source'')
{\tt e.act(r,v)}, presents {\tt (s, adjoint(v))}, ({\tt s}=source of {\tt v})
to be acted on by the next enzyme following {\tt e} in an enzyme chain.
Predicates may make the presentation conditional. 

There are also quantors such as {\tt fALL}
 which invokes the action
 {\tt e.act(r,v)} of an enzyme {\tt e} for all valences {\tt v} of a 
radical {\tt r} which fulfill a condition; {\tt e} is appended as tail of 
an enzyme chain with head {\tt fALL}. ( A {\em forAll} Quantor exists also for 
predicates.)     

Unless one  uses the concurrent version of the dynamics (cf. section 
\ref{sec:Concurrency}),
 the present implementation for single processor machines
offers the option of reporting an enzyme to an agenda together with the
argument pair {\tt (r,v)} it will act on,
 instead of attaching it to the valence {\tt v}
or the object of {\tt r} in the \system . There are enzymes to do that. 
A $\_ Sag$-enzyme (``source-to-agenda'') in an enzyme chain reports
 the whole chain together 
with {\tt (s, NULL)}, {\tt s} the source of its {\tt v}-argument. 

{\em Membranes}: The ``push membrane'' enzymes $\_ PIM$ and $\_ POM$ 
surrounds a radical by a membrane (attached to all its valences) or pushes
 the radical inside resp. outside a membrane that goes through one of its valences resp. its adjoint. The membrane's code is specified by a predicate. 

{\em Paths}: There are enzymes to make or prolong paths along specified 
valences.

\section{Graphic Interface}
There is a graphic interface which can be used to construct \system s
and manipulate them by use of enzymes. All this is done by mouse click, except
that numerical or text-data need to be entered from the key board when they
are not copied from somewhere.

All operations are recorded and translated into code by use of the LISP-like
scripting language which is described in the next section. In this way,
programs which construct \system s or code for processes 
on \system s can be composed by mousclick. 

In detail one may\\
- Position objects or delete them\\
- Link them by valences (with or without adjoint link) or delete links.\\
- Enter data (\emph{Value}s) into objects or links.\\
- Select basic enzymes and accompanying predicates from lists.\\
- Build composite enzymes as chains of basic ``micro-enzymes''.\\
- Invoke conditional application of enzymes to radicals {\tt r} with or without a selected valence {\tt v}.\\
Among the enzymes there are some which \\
- create a membrane around an object, or push a membrane beyond other objects.\\
- create paths or prolong them\\
- create copies of individual objects, of the whole system, or of a subsystem bounded by a membrane.\\
- induce sweeps through the whole \system \ [or through a part of it which is bounded by a membrane or 
marked by path-Links], applying the tail of an $aFRK$-enzyme to every radical in it once.\\
- induce sweeps which are reflected at the locus of maximal path distance, assemble data from all the objects and links of the system and report them to the \system ' root. The computation of maxima of real data at all objects is a sample application.\\
- induce diffusion within the \system \  or a subsystem.\\

Coordinates are assigned to each object. A panel allows to rotate and translate the coordinate system. Different windows admit different views on a system and its dynamical behavior.

\subsection{Permanence: XML}
Methods are provided to translate a \system \ into  {\bf XML} and conversely
to construct a \system \ from {\bf XML}-code. The methods can be invoked 
through the graphic interface. 

In this subsection we assume the reader is familiar with {\bf XML}. An
introduction to {\bf XML} can be found in \cite{XML}.

The dtd ({\it document type definition}) listed in \cite{JW} defines tags and
their possible attributes for the systems elements. An internal id number is assigned to each object
and valence to let a valence remember its source, target and adjoint valence.
\\

The {\tt <system>} tag encloses anything else inside the document.
The {\tt <object>} tag carries an objects id number and a position as attributes. {\tt <value>} tags may be enclosed.
The {\tt <valence>} tag carries a valences id number, the id numbers of its source and target objects and of its adjoint valence as attributes.
Again, {\tt <value>} tags may be enclosed.
The {\tt <value>} tag carries the value type and the value content as attributes.\\
\\
The way XML is designed it allows future extensions
while assuring up- and downward compatibility.
\\

\section{The Scripting Language}

\begin{figure}[h!]
  \begin{center}
    \leavevmode
     \input{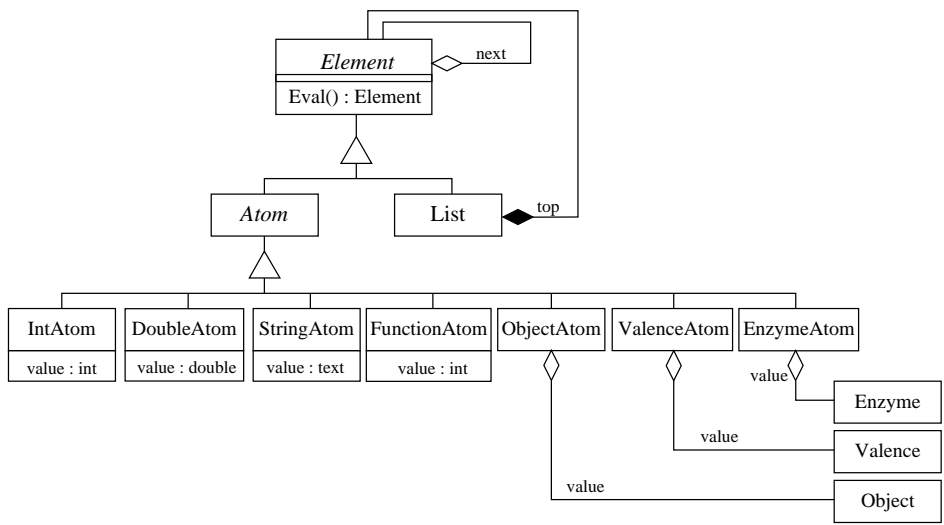}
  \end{center}
  \caption{\small The class diagram for the SyCL parser.}
  \label{parser}
\end{figure}

If a particular kind of complex problem occurs very often, it might be worthwhile to express instances of the problem
as sentences in a simple language\cite{GHJV}. Solving a problem then means to interpret the sentence, for which an {\it interpreter}
is needed.\\
Based on a certain grammar which defines a simple language, the interpreter represents and interprets sentences in that
language. To be efficient, character strings which make up the sentences are transformed into trees of objects, which are easy to
handle by the interpreter. This transformation is done by a software called {\it parser}.\\
The grammar used for our purposes is very similar to the one used by the {\it list processing language LISP}.
\begin{verbatim}
expression ::= list | atom
list ::= `(' expression* `)'
atom ::= `a' | `b' | ... | `z' | `+' | `-' | `*' | `/' | `$' | 
             { `a' | `b' | ... | `z' }* | ` quote ' expression
\end{verbatim}
In object oriented software design, a well known interpreter pattern (suggested in ref.\cite{GHJV}) has shown to work fine in most cases.
To implement a parser and interpreter for the SyCL program package, this pattern is used in combination with a composite
pattern \cite{GHJV} and extended by one more abstraction in the way shown in figure \ref{parser}.\\
$\;$\\
The \textbf{\textit{Element}} class is abstract. It provides polymorphism by assuring that every object is of type {\it Element}.
A Element can thus either be an object of type {\bf List}, which consists of further Elements, or an object of type
\textbf{\textit{Atom}}, where {\it Atom} is abstract again. The following classes are derived from class {\it Atom}.\\
$\;$\\
A list can consist of elements of any kind. To evaluate a list, the first element has to be a function, or an enzyme,
where the following elements serve as parameters, which have to be evaluable unless they are quoted (i.e. they
start with a quote).\\

\subsection*{\it Implemented Functions}

The commands, this interpreter can execute are listed below. Enzymes can be
applied as well, taking a radical or a list
of a radical and a valence as a parameter.\\

\noindent
{\bf (adjoint v)} returns the adjoint valence to valence {\it v}\\
{\bf (append expr li)} appends the expression {\it expr} to the list {\it li} and returns this list. The original list remains unchanged!\\
{\bf (car li)} returns the first element of the list {\it li}\\
{\bf (cdr li)} returns the list {\it li} without its first element\\
 {\bf (connectto a x y z)} searches the system for an object at the position $(x,y,z)$ and assignes the variable {\it a} to it.\\
{\bf (delete a)} removes element {\it a} from the variable list and deletes it (compare {\it remove}).\\
{\bf (edit a b)} changes the value of an object or valence {\it a} to the SyCL value {\it b}.\\
{\bf (enzyme expr)} defines a new enzyme from the expression {\it expr}. {\it expr} has to be of the form {\it xxxx=enz1enz2enz3....} and
must be quoted in order not to be executed as a command! The newly defined enzyme is automatically inserted into the enzymelist as well as the enzyme menu from the GUI.\\
{\bf (enzymes)} returns the list of enzymes.\\
{\bf (eval expr)} evaluates the expression {\it expr}.\\
{\bf (equal a b) or (eq a b)} returns 1 if {\it a} and {\it b} are equal, an empty list otherwise.\\
{\bf (for (n i0 i1) ( expr1 expr2 expr 3 ... ))} evaluates the expressions {\it expr1, expr2} and {\it expr3 ...}
        with the variable {\it n} taking integer values from {\it i0} to {\it i1}\\
{\bf (funlist)} returns the list of implemented functions.\\
{\bf (greaterthan a b) or (gth a b)} returns 1 if {\it a} is greater than {\it b}, an empty list otherwise.\\
{\bf (if expr ( expr1 expr2 expr3 ... ))} evaluates the expressions {\it expr1, expr2} and {\it expr3 ...} if {\it expr} is true.\\
{\bf (length li)} returns the number of elements of the list {\it list}.\\
{\bf (lessthan a b) or (lth a b)} returns 1 if {\it a} is less than {\it b}, an empty list otherwise.\\
{\bf (list a b ... )} returns a list with the parameters {\it a, b, ...} als elements.\\
{\bf (load filename)} loads and evaluates commands in a file {\it filename}. This can be done more comforably with a file-select box via the GUI {\it open file} menu entry.\\
{\bf (minus a b) (- a b)} returns the difference of {\it a} and {it b}.\\
{\bf (newo a)} creates a new SyCL object of SyCL value {\it a} and returns an object element.\\
{\bf (notequal a b) or (neq a b)} returns an empty list if {\it a} and {\it b} are equal, 1 otherwise.\\
{\bf (nth n li)} returns the {\it n}th element of a list {\it li}. Counting starts at zero!\\
{\bf (nthval n a)} returns the {\it n}th valence of the valences directed toward the object {\it a} (counting from 0).\\
{\bf (oblist)} returns the list of variables.\\
{
{\bf (poso a x y z)} repositions the object designated by variable {\it a} at $(x,y,z)$.}
{\bf (plus a b) (+ a b)} returns the sum of {\it a} and {\it b}.\\
{\bf (rand n)} returns an integer random number between zero (inclusive) and {\it n} (exclusive).\\
{\bf (remove a)} removes element {\it a} from variable list (compare {\it delete}).\\
{\bf (replacenth n li expr)} replaces the {\it n}th element of a list {\it li} by {\it expr} and returns this list. The original list {\it li} remains unchanged!\\
{\bf (run '(e v))} lets the enzyme {\it e} act on the source of valence {\it v}.\\
{\bf (set a b)} same as {\bf setq}, but evaluates {\it a} first.\\
{\bf (setl l v a b)} {\it a} and {\it b} have to be object elements. An unidirectional valence is created between {\it a} and {\it b}
        with a SyCL values {\it v}. The valence element is assigned to the variable {\it l}.\\
{\bf (seto a ['int $|$ 'real $|$ 'bool $|$ 'enz] ['vector m $|$ 'covector n $|$ 'matrix m n] b... [x [y [z]]])} creates a SyCL object with a SyCL value {\it b} (evaluated) and assigns the object element to the variable {\it a}. Creating a vector of dimension m, a covector of dimension n, or a matrix of dimension {\it m}x{\it n} requires {\it m}x{\it n} elements {\it b}! {\it x, y, z} can be appended to specify coordinates.\\
{\bf (setq a b)} creates an expression, sets it to {\it b} (evaluated) and assigns it to the variable {\it a}.\\
{\bf (setv v v0 v1 a b)} {\it a} and {\it b} have to be object elements. A bidirectional valence is created between {\it a} and {\it b}
        with SyCL values {\it v0} and {\it v1} for the valence and its adjoint. The valence element is assigned to the variable {\it v}.\\
{\bf (times a b) or (* a b)} returns the product of {\it a} and {\it b}.\\
{\bf (valences a)} returns the number of valences directed toward an object {\it a}.\\

The following script may serve as an example of motion along a valence $v$ between two objects $a$ and $c$:
\begin{verbatim}
     (seto a 1)
     (seto b 2 100)
     (seto c 3 0 100)
     (setv v 1 1 a b)
     (setq w (adjoint v))
     (setv m 1 1 a c)
     (setq q (adjoint m))
     (enzyme 'move=_AMR_RMV{?nAD}_PRS_RFU)

     (move '(a m))
\end{verbatim}
The last command invokes the action of the above defined enzyme {\tt move} with object $a$ and valence $m$ as parameters.

\section{Some basic Processes}
\subsection{Enzymatic Copying}
\label{sec:splitFork}
Let us call a link {\em bidirectonal} if its adjoint is also a link, 
{\em unidirectional} otherwise. Note that the terms are used to characterize 
individual links, not pairs (link, adjoint link). 

There exists an enzyme, call it the splitFork-enzyme, whose continued 
conditional application to a finite \system \ $\S $ whose links are all 
bidirectional, produces after some finite time two copies of $\S$.

\begin{figure}
\begin{center}
\setlength{\unitlength}{0.001875in}%
\begin{picture}(1474,180)(123,460)
\thicklines
\put(1310,470){\circle*{14}}
\put(1310,630){\circle*{14}}
\put(1160,640){\vector( 1, 0){140}}
\put(1000,460){\vector( 1, 0){140}}
\put(1150,470){\circle*{14}}
\put(1150,630){\circle*{14}}
\put(1000,640){\vector( 1, 0){140}}
\put(1140,620){\vector(-1, 0){140}}
\put(1140,480){\vector(-1, 0){140}}
\put(1300,620){\vector(-1, 0){140}}
\put(1580,540){\vector(-1, 0){140}}
\put(1440,560){\vector( 1, 0){140}}
\put(860,540){\makebox(0,0)[b]{\smash{$\Longrightarrow$}}}
\put(990,470){\circle*{14}}
\put(1300,480){\vector(-1, 0){140}}
\put(1160,460){\vector( 1, 0){140}}
\put(990,630){\circle*{14}}
\put(1590,550){\circle*{14}}
\put(1420,540){\vector(-3,-2){ 96.923}}
\put(130,470){\circle*{14}}
\put(140,640){\vector( 1, 0){140}}
\put(280,620){\vector(-1, 0){140}}
\put(300,620){\vector(3, -2){ 96.923}}
\put(410,550){\circle*{14}}
\put(130,630){\circle*{14}}
\put(290,470){\circle*{14}}
\put(290,630){\circle*{14}}
\put(570,550){\circle*{14}}
\put(730,550){\circle*{14}}
\put(400,540){\vector(-3,-2){ 96.923}}
\put(560,540){\vector(-1, 0){140}}
\put(580,560){\vector( 1, 0){140}}
\put(1430,550){\circle*{14}}
\put(1320,620){\vector(3, -2){ 96.923}}
\put(720,540){\vector(-1, 0){140}}
\put(280,480){\vector(-1, 0){140}}
\put(140,460){\vector( 1, 0){140}}
\put(420,560){\vector( 1, 0){140}}
\end{picture}
\end{center}
 \caption{Replication fork dynamics.}\label{fig:splitFork}
 \end{figure}
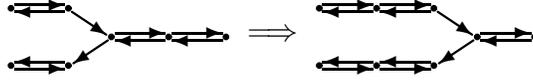

This is a generalization of the asymmetrical replication fork mechanism 
which the living cell uses to copy DNA \cite{Alberts}. 
It works not only for 
chains of pairs of bidirectional links (which mimick the 
double helix of DNA), but for any \system \ whose
links are all  bidirectional, independent of its topology.
[All the interna like (possibly overlapping) internal structure of constituent objects, membranes and pathLinks marking various subsystems are copied as 
well]. 
The operation on chains is shown in figure \ref{fig:splitFork}
 The
description of the action $s_X$ of the splitFork-enzyme
 to the radical of object $X$ 
is as follows. 

\begin{figure}[h!]
  \begin{center}
    \leavevmode
     \input{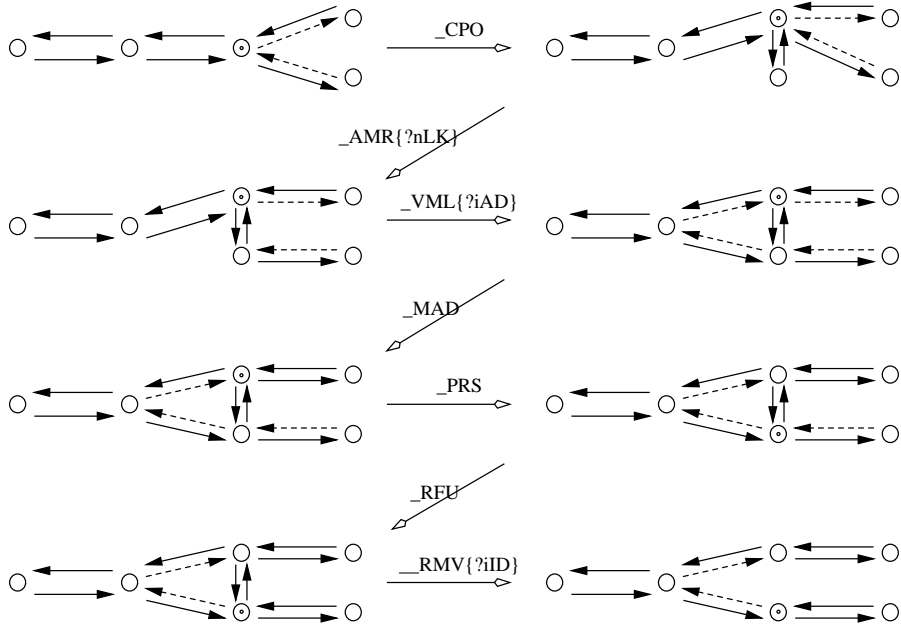}
  \end{center}
  \caption{\small One step within a sweep of the SplitFork enzyme: {\tt sFRK = \_CPO \_AMR\{?nLK\} \_VML\{?iAD\} \_MAD \_PRS \_RFU \_RMV\{?iID\} \_Sag\{?nAD\}}}
  \label{fig:microFork}
\end{figure}

\begin{enumerate}
\item A copy $X^\prime $ of $X$ is made.
\item The links incident on $X$ other than loops are distributed among $X$ and its copy as follows:
\begin{itemize}
\item[-] bidirectional links with target $X$ get $X^\prime$ as their target
\item[-] unidirectional links with target $X$ retain $X$ as their target
\item[-] bidirectional links with source $X$ retain $X$ as their source
\item[-] unidirectional links with source $X$ get $X^\prime $ as their source
\end{itemize}
The loops $X \mapsto X$ remain in place
 and get a copy $X^\prime \mapsto X^\prime$.
\item The adjoints of formerly unidirectional links are promoted to the status
of links. 
\end{enumerate}
One may imagine that the copy $X^\prime$ is connected by a pair
of adjoint bidirectional identity links to $X$ to begin with. Then the management of the links can be interpreted as instances of motion - multiplication 
with identity links and creation of fundamental adjoints. 
In the end, the identity links between $X$ and $X^\prime $ are killed. 
This decomposition of the splitfork-enzyme into an action of 
micro-enzymes is illustrated in figure \ref{fig:microFork}.

Define a fork at $X$ as a pair of links, both unidirectional, one with 
target $X$ and the other with source $X$. 

\begin{theorem}{\em (Universal copy constructor)} \label{theo:cpy}
Let $\S_0$ be obtained from a finite connected system $\S$ whose links are 
all bidirectional
by action of $s_{X_0}$ at some $X_0\in \S$. For $t>0$, let $\S_t $
be obtained from $\S_{t-1}$ by action of $s_X$ for all objects $X$ such that 
there is a fork at $X$.

$\S_t$ is well defined for $t\geq 0$. For 
sufficiently large $t$, it is independent of $t$ and
consists of two disconnected copies of $\S$.
\end{theorem}
``Once replication has started, it continues until the entire \system \ has 
been duplicated''. Upon substituting ``genome'' for ``\system `` this becomes 
a quote from a genetics text book \cite{genes}.  

This theorem was first demonstrated in \cite{mack:kyoto}.

The fact that $s_X$ is mathematically well defined  
is somewhat subtle. It rests on a theorem, proven in \cite{mack:cmp},
that a \system \ can be specified up to isomorphism by 
enumerating its arrows, which of them can be composed and which arrow is the
 result. One must also indicate which arrows are links and what are their 
adjoints. But nothing need be said about objects - they can be reconstructed. 
Using this, the above theorem can be proven, with the understanding that 
the phrase ``two copies of $\S$'' means ``two \system s, both isomorphic to
$\S$''. The isomorphism class of $\S $ does not retain the 
information about the internal structure of non-atomic objects
 (\emph{black boxes}). But this information can be retained by the copies.
To retain it, one uses the universal copy constructor to copy objects 
of $\S$ which are themselves \system s, to copy their non-atomic constituents,
and so on. This does not continue {\em ad infinitum} by the 
axiom of non-self-inclusion.

The fact that the dynamics is well defined rests on the fact that 
$s_Xs_Y=s_Ys_X$ for the kind of \system s that occur as $\S_t$. 

The action of the splitFork-enzyme is quite robust against 
errors due to computer failures which mimic local mutations.
But there is one exception which leads to catastrophic results of the type of 
Down's syndrome - a third copy is made of part of the \system . This happens 
when a fundamental adjoint gets lost (or added) ``at the wrong moment'',
cf. \cite{mack:kyoto}. 
\footnote{In man, Down's syndrome is caused by the presence of three
copies of chromosome no. 21 instead of the usual two}

There is a more flexible 
(but not quite as axiomatic) version of the splitFork-enzyme which can be used
to restrict copying to subsystems bounded by membranes. It employs 
membranes in place of the prototypical membrane made by unidirectional 
links.

\subsection{Digestion}
In animals, building blocks of newly made structures must first be 
prepared by degrading ingested organic material by metabolic processes. 

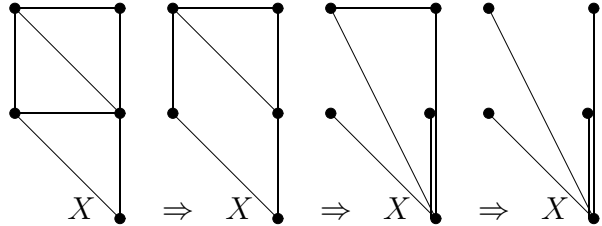
\begin{figure}
\label{fig:digest}
\begin{center}
\setlength{\unitlength}{0.7mm}
\begin{picture}(40,50)(40,0) 
\put(0,0){\begin{picture}(0,0)
\put(20,0){\circle*{2}}
\put(0,20){\circle*{2}}
\put(20,20){\circle*{2}}
\put(0,40){\circle*{2}}
\put(20,40){\circle*{2}}

\put(20,0){\line(-1,1){20}}
\put(0,20){\line(1,0){20}}
\put(20,20){\line(-1,1){20}} 
\put(0,20){\line(0,1){20}}
\put(20,0){\line(0,1){40}}
\put(0,40){\line(1,0){20}}
\put(10,0){$X$}
\put(28,0){$\Rightarrow$}
\end{picture}}

\put(30,0){\begin{picture}(0,0)
\put(20,0){\circle*{2}}
\put(0,20){\circle*{2}}
\put(20,20){\circle*{2}}
\put(0,40){\circle*{2}}
\put(20,40){\circle*{2}}

\put(20,0){\line(-1,1){20}}
\put(20,20){\line(-1,1){20}} 
\put(0,20){\line(0,1){20}}
\put(20,0){\line(0,1){40}}
\put(0,40){\line(1,0){20}}
\put(10,0){$X$}
\put(28,0){$\Rightarrow$}
\end{picture}}

\put(60,0){\begin{picture}(0,0)
\put(20,0){\circle*{2}}
\put(0,20){\circle*{2}}
\put(18.8,20){\circle*{2}}
\put(0,40){\circle*{2}}
\put(20,40){\circle*{2}}

\put(20,0){\line(-1,1){20}}
\put(20,0){\line(0,1){40}}
\put(19,0){\line(0,1){20}}
\put(20,0){\line(-1,2){20}}
\put(0,40){\line(1,0){20}}
\put(10,0){$X$}
\put(28,0){$\Rightarrow$}
\end{picture}}

\put(90,0){\begin{picture}(0,0)
\put(20,0){\circle*{2}}
\put(0,20){\circle*{2}}
\put(18.8,20){\circle*{2}}
\put(0,40){\circle*{2}}
\put(20,40){\circle*{2}}

\put(20,0){\line(-1,1){20}}
\put(20,0){\line(0,1){40}}
\put(19,0){\line(0,1){20}}
\put(20,0){\line(-1,2){20}}
\put(10,0){$X$}
\end{picture}}

\end{picture} 
\end{center}
\caption{Digestion enzyme attacks at $X$}
\end{figure}

Here we describe an enzyme whose continues action $d_X$ at an object $X$
of a connected \system \ degrades its internal structure in the manner
 shown in figure \ref{fig:digest}.

$d_X$ consists of consecutive steps.

1. (Death) The far side of all triangles of 3 links with tip $X$ is removed, 
together with  their adjoints. 

2. (Motion)
 If $b$ is a link from $Y\neq X$ to $X$ and $b^\prime $ is a link from $
Y^\prime $ to $Y$ then $b\circ b^\prime $ becomes a link and $b^\prime $
ceases to exist as a link. 

3. Fundamental loops $X\mapsto X$ are removed.

Actually the 3rd step can be dispensed with when step 2 operates also for 
$Y=X$. 

\label{subsec:digestion}
%

\subsection{Reaction Diffusion}
Restrict attention to those values of objects which can be added
and subtracted.

Consider a radical {\tt r} with object $X$ and its valences {\tt v}
with source objects $Y_v$. There is an enzyme which does the following. 

For every valence {\tt v} with source $Y_{{\tt v}}$
(or those which fulfill the condition set by a predicate)
 it takes a fraction $\alpha \in {\bf R}$ of the 
original value $\xi $ of $X$, subtracts it from $\xi$, parallel transports it 
to $Y_{{\tt v}}$ along {\tt v} and adds the result to the value
 $\eta_{{\tt v}}$ of $Y_{{\tt v}}$. Conversely, it takes
a fraction $\alpha$ of the original value $\eta_{{\tt v}}$, 
subtracts it from $\eta_{{\tt v}}$, parallel 
transports it to $X$ and adds the result to $\xi$. In the most important 
examples, the parallel transport is trivial (i.e. $\xi $, $\eta $ are 
invariants). This models diffusion. 

One may add loops $X\mapsto X$ which effect nonlinear maps of the values
of $X$ as ``parallel transports''. This models reactions, e.g. chemical 
reactions. 

\subsection{Shock fronts}
\label{sec:adaptFork}
The splitFork dynamics is an example of the mechanism of propagating 
shock fronts. It is an abstraction of the most essential feature 
of shock fronts in nature.
\footnote{Shock fronts  occur  in nature 
as a consequence of nonlinearity in excitable media
 \cite{erregbareMedien}
 when the passage 
of an excitatory event is followed by a dead time.
It has the consequence that the excitation can only propagate in the
outward direction because  a renewed 
excitation is prohibited during the dead time. 
Forest fires are a well known and much studied 
example \cite{forestFire}. Natural neurons also have a dead time after firing. 
}
 In the splitFork dynamics, there is at any time 
a bipartite boundary between the part of the \system \ which has not 
yet been copied, and the two copies of the rest. The two boundaries with 
the two copies have opposite orientation. They are made of the 
unidirectional fork prongs. The excitation ($sFRK$-action) 
is restricted to radicals which have forks, and
it can only pass through 
bidirectional links into the part of the system which has not been copied yet. 

Similarly, a single boundary made of unidirectional links  may be made to
 propagate unidirectionally.
 There is an enzyme $aFRK$ to do that.
  It generates sweeps through the system. 
Making an enzyme chain with head $aFRK$ and some tail will make the tail act   
on every radical {\tt r} of the \system .
 This may be used to induce relaxation sweeps, for instance.

\subsection{Cellular growth}

\begin{figure}[h!]
  \begin{center}
    \leavevmode
     \input{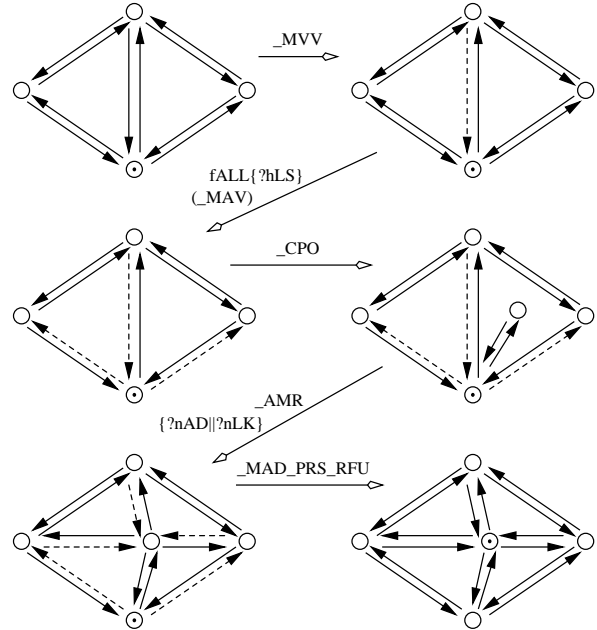}
  \end{center}
  \caption{\small Cell replication in a system theoretic diagram: {\tt cell = \_MVV fALL \{?hLS\}(\_MAV) \_CPO \_AMR \{?nAD$||$?nLK\} \_MAD \_PRS \_RFU }}
  \label{cellenzyme}
\end{figure}

Cell replication can be achieved by realization of the following idea: A cell $A$ gets triggered to replicate, for instance by a certain
value of a gradient to a neighboring cell $B$. A copy $C$ of $A$ is made and placed between $A$ and $B$ in order to smoothen the gradient.
Therefore the fundamental links between $A$ and $B$ are replaced by new fundamental links between $B$ and $C$. Furthermore $C$ obtains
new fundamental links to all neighbors, $A$ and $B$ are both connected to (see figure \ref{cellenzyme}).\\

\subsection{Assembly of data}
\label{sec:backwardFork}
Consider a finite \system \
with a distinguished radical {\tt root}.
We describe another shock wave mechanism, call it assemblyFork,  
which can be used to assemble data from the system at {\tt root}. 

 First one sends out a shock wave
which builds a spanning tree - details are given below. 

Take any data which are of type {\em AlgebraicValue}. They may be 
attached to objects or links.
The assemblyFork-dynamics computes the sum of these algebraic values, parallel transported along the branches of the tree, and deposits it at {\tt root}.

 For instance, the data maybe sets of strings
\footnote{sets become lists by lexicographic ordering}, 
as occur in encodings of data in {\bf XML},
where $\circ $ is concatenation of strings, and $\oplus$ is disjoint union of 
sets. No loss of information is involved in taking the sum (disjoint union) 
in this case. 

Or the data may be the (real) elements of a {\em max-plus}-algebra, where 
addition amounts to taking the maximum, and multiplication is addition of
real numbers. The maximum of all the data is computed in this case.  

If the data are invariants (cf. earlier) then, by definition, it would make no
difference if parallel transport were along any  path to {\tt root} 
other than the path  along the tree.
\footnote{More generally, if all links are unitary, i.e. if $b^\ast = b^{-1}$,
the results differ at most  by a gauge transformation, cf. \cite{mack:cmp}.}  

When a shock proceeds up along any of the branches of the tree, 
a leaf will eventually be reached where no prolongation of the branch is
possible without creating a loop. Then the shock gets reflected
 and propagates down 
the tree, taking data from the leaf along via 
parallel transport (cf. section \ref{sec:algebraicValue}).
 At any node of the tree, 
the shock has to wait until the 
reflected shocks from all the branches are in. Then the data which they 
transport are all added up and the result is transported further down the
tree. 

The {\em max-plus} version of this mechanism is 
important when one does optimization by relaxation. One needs 
to know the maximum of a measure of the local error in order to know 
when to stop. 

Let us finally describe how the spanning tree is made and marked. 
In principle, the order of the valences in the valence list of a radical 
is irrelevant. But we may either make an exception for the purpose of
the assemblyFork-mechanism, distinguishing the first valence as a 
``principal port''. Or we may use  paths to mark the branches of the tree. 
Suppose we do the first; generalization will be  obvious.

Given a distinguished radical {\em root}, the \system \ decomposes into shells
labeled by path-distance
\footnote{path distance was defined in subsection \ref{sec:basicData}}
 $n \geq 0$ from {\tt root}, and links between
 adjacent shells. The shells are like the shells of an onion. 
 The $n$-th shell is a subsystem which contains all objects 
with path distance $n$ from {\em root}, and the links between them. 

A shock emanating from {\tt root} propagates from shell to shell in 
order of increasing $n$.

When starting at shell 0, the tree has no link.

 When the shock reaches shell $n$, there is a membrane 
= boundary between shell $n$ and $n+1$, and the tree  $n$ will 
have been constructed up to shell $n$.
 Now one selects in some arbitrary way links from 
the boundary such that each object $X$  of the $n+1$st shell is source 
of exactly one such link. This can be done by local operations. 
 The adjoint of the unique link 
with source $X$ is made principal port of $X$. In this way the tree has 
been grown to shell $n+1$. Now one continues. In a finite \system , there
are finitely many shells. So the process comes to an end.

\section{Outlook}
There exists also an (exterior) differential calculus and geometry on 
\system s \cite{Dimakis,mack:cmp}.
 It is intermediate between standard exterior differential calculus
and geometry  on manifolds and non-commutative differential calculus and
geometry \cite{Connes}.
 It satisfies $d^2=0$ and the Leibniz rule, the algebra of functions 
is commutative, and derivatives of functions are ordinary finite difference
derivatives. It has not been implemented in software yet, but we plan 
to do so  in the future.

Another line of current investigation is multiscale analysis \cite{MGrid}. 
By definition, genuinely complex systems show ``emergent''
properties that are not shared by their subsystems with few objects. 
The basic idea of multiscale analysis, including multigrid
methods\cite{MGrid} and the renormalization group \cite{RG},
 is that although a genuinely complex 
system can not be understood as a whole by studying reasonably small
subsystems in isolation, a complexity reduction is often possible by doing so.
To this end one introduces new objects which represent subsystems, but 
one retains only that part of the information on their internal structure  
(including functionality, i.e. enzymes) that is relevant for the cooperation
that causes emergent - i.e. nonlocal - phenomena. A related plan is 
known in information science under the name of {\em integration}
\cite{EhrigOrejas}. Some theoretical considerations concerning this 
were presented in \cite{mack:cmp} and in \cite{tomkos}.
Relaxation sweeps can be implemented by enzymatic computation, and it is 
straightforward to extend them to \system s which have a multilevel
structure like multigrids. The challenge is to extend the domain of
applicability of existing multigrid and renormalization group technology,
including in particular disordered systems.

Finally it is a challenge to invent an ``enzymatic game of life'', and 
to implement within the present frame as much of biochemical and larger
scale processes that constitute life. This would be in the spirit
of work on Artificial life \cite{artificialLife}.
\subsection{Towards an infome project}
A much more challenging and long range plan 
would be an {\em infome} project which amounts
to mapping into software what is known about structure and function
in the living cell in such a way that processes that constitute life can be 
simulated {\em in silico}. After the genome project and the 
envisaged proteome project \cite{proteome} 
 which is supposed to classify all proteins that
a human body makes in its lifetime and their function, this is a logical next
 step. 

Work on immunology along similar lines is in progress \cite{schelli};
the establishment of a data base has been the proposed. 
For a discussion of the cognition problems that are of interest 
in immunology see \cite{immuno}. 

\section{Acknowledgement} 
We thank the Deutsche Forschungsgemeinschaft for financial support,
and the German Israel foundation for a travel grant. We would like to express 
our thanks to Achi Brandt, Irun Cohen, Martin Meier-Schellersheim, Dirk Rathje,
 Sorin Solomon, and York Xylander for numerous enlightening discussions. We also thank Daniel L\"ubbert, Bleicke Holm, Marcus Speh and York Xylander for their collaboration in earlier stages of this project.

\end{document}